\documentclass[%
 article,
nofootinbib,
 amsmath,amssymb,
 aps,
showkeys,
]{revtex4-2}

\usepackage{graphicx}
\usepackage{dcolumn}
\usepackage{bm}
\usepackage{color,hyperref}
\hypersetup{colorlinks,breaklinks,
            linkcolor=blue,urlcolor=blue,
            anchorcolor=blue,citecolor=blue}
\usepackage{xurl}
\usepackage{color,soul}

\begin{document}

\newcommand{\bbe}{\begin{eqnarray}}
\newcommand{\ee}{\end{eqnarray}}
\newcommand{\n}{\nonumber}



{\Large \bf \raggedright Moderate Correlation between the Accretion Disk and Jet Power in a Large Sample of Fermi Blazars}

\vspace{1.5mm}

{\raggedright Garima Rajguru and Ritaban Chatterjee}

{\raggedright Department of Physics, Presidency University, 86/1 College Street, Kolkata-700073, WB, India.}

\maketitle

{\small \raggedright We present the results of studying the accretion disk vs jet power for a large fraction of all the blazars detected by the Fermi Gamma-Ray Space Telescope. The disk power is inferred from the emission line luminosities obtained from published results. As indicators of jet power, we use low frequency radio luminosity from the extended jet, maximum speed of radio knots observed in the VLBA monitoring of the pc-scale jets, kinetic energy of electrons in the jet deduced from the best-fit theoretical models of their spectral energy distribution, and $\gamma$-ray luminosity with and without beaming correction. We obtain a significant correlation in most of those cases. However, we find that the correlations are often driven by the common redshift dependence of the compared quantities. In order to remove the redshift bias and probe the intrinsic correlation between the disk and jet power, we compute the partial correlation coefficient as well as the correlation in small redshift bins, and find that the intrinsic disk-jet correlation is still present but weaker. In the cases, in which the common redshift dependence does not affect the result, we find that blazars do not exhibit high jet power for low disk luminosities while there are both high and low jet power for high disk luminosities. This result indicates that a powerful disk is a necessary but not sufficient condition to produce a powerful jet.
{\bf Comments:} Published in Physical Review D in
the 15 September 2022 issue (Vol. 106,
No. 6).\\
DOI: \url{https://doi.org/10.1103/PhysRevD.106.063001}}


\section{Introduction}

According to the unified scheme of active galactic nuclei (AGNs), observed properties of various types of AGNs are linked to their orientation with our line of sight \citep{1995PASP..107..803U}. One class of AGNs, which possess energetic jets pointed almost directly towards the Earth with an angle $\theta <10^\circ$, is called blazars. 
A strong radio signal is considered to be characteristic of the jet \citep{1995PASP..107..803U}, arising from synchrotron radiation from the relativistic electrons present in it \citep{1981Natur.293..714B,1982ApJ...253...38U,1988AJ.....95..307I,1998ASPC..144...25M}. The synchrotron spectrum extends from radio to optical (sometimes X-ray) frequencies. The high energy (keV$-$GeV) emission is presumably due to the inverse-Compton effect, which up-scatters the same jet synchrotron photons \citep{1992ApJ...397L...5M, 2002ApJ...564...92C, 2005ApJ...627...62A} or external photons, from the broad line region (BLR) or dusty torus, to X-rays and $\gamma$-rays \citep{1994ApJ...421..153S, 1999ApJ...521L..33C, 2000ApJ...545..107B, 2009ApJ...692...32D}. The accretion disk emits a multi-color blackbody spectrum \citep{1973A&A....24..337S,malkan&sargent1982} with a peak in the optical-UV wavelengths. 

Blazars may be categorized into two major groups based on the equivalent width (EW) of their broad emission lines, in the rest frame. BL Lacertae (BL Lac) objects are a type of blazar, which are supposed to be devoid of prominent emission or absorption lines (EW $<5$\AA), whereas flat spectrum radio quasars (FSRQ) possess broad emission lines (EW $>5$\AA) in their spectra \citep{1995PASP..107..803U}. However, this segregation may not be robust \citep{2011MNRAS.414.2674G}, and BL Lacs have been shown to exhibit emission lines with EW
 $>5$\AA~ as more precise observations have been available in recent years \citep{1995ApJ...452L...5V,2010A&A...516A..59C}.

The mechanism of the launching of the jet is not well known. Widely accepted theories include the extraction of rotational energy of the super-massive black hole (SMBH) at the center \citep{1977MNRAS.179..433B} or extraction of power from the accretion disk in the presence of magnetic fields sustained by the disk \citep{1982MNRAS.199..883B}. While the details are not known the mechanism of jet launching and collimation is assumed to be related to the accreting matter and magnetic fields associated with the disk \citep{lind&meier1989,mckinney&narayan2007,Tchekhovskoy&narayan2011,2013ApJ...770...31W}. Thus, a relation between the dynamics of the disk and the jet, and its manifestation in the disk and jet power are anticipated. 

One of the first attempts to probe the accretion disk-jet relation compared the jet power of a sample of radio galaxies to the luminosity of the narrow emission lines \cite{1991Natur.349..138R}.
The broad and narrow emission lines are produced in gas clouds photoionized by the continuum emission of the accretion disk \citep{1968ApJ...151...71W,1968ApJ...151..807O}. Therefore, the luminosity of the broad and narrow emission lines are often used as an estimate of the disk luminosity. For example, 10\% of disk luminosity ($L_{disk}$) is assumed to be reprocessed as the luminosity of the BLR ($L_{BLR}$). The disk luminosity, as estimated from $L_{BLR}$, \textit{versus} the jet power from resolved very long baseline interferometry (VLBI) observations of a sample of 55 blazars, was investigated by \cite{1997MNRAS.286..415C}, who found a weak correlation.

Similar work followed with different samples of blazars and different measures for jet power. An approximate equality between jet power obtained from SED modeling of multi-wavelength data, and disk luminosity inferred from $L_{BLR}$ was obtained by \cite{2003ApJ...593..667M,2008MNRAS.391.1981M,2009MNRAS.399.2041G}. VLBI measurements \cite{2004ApJ...615L...9W} and SED modeling \cite{2014Natur.515..376G} was used to estimate the jet power of blazars, and their comparison with disk luminosity showed a clear correlation. A considerable connection of the disk and jet power in blazars has been shown in \cite{2014MNRAS.445...81S,2010MNRAS.402..497G}. A sample containing 225 FSRQs, 42 BL Lacs and radio galaxies has been used by \cite{2014MNRAS.445...81S}, who estimated their jet power from $\gamma$-ray and radio luminosity. 
A significant trend in the disk and jet power in \textit{Fermi} 2LAC blazars has been found by \cite{2016NewA...46....9D}. They collected Doppler factors from the literature to correct the observed GeV luminosity for relativistic beaming. They found that the trend followed by the FSRQs was different from that of the BL Lacs. Using radio core observations, \cite{2019Ap&SS.364..123C} found that FSRQs generally have larger accretion disk luminosity than the jet radiation power. 

There is a consensus in estimating the disk luminosity from $L_{BLR}$ while tracing the jet luminosity remains elusive. This is because the jet carries both radiative and kinetic power in unknown proportions. Moreover, in the case of blazars, the jet is Doppler beamed toward us due to small angles with our line of sight. The value of the beaming factor is difficult and time intensive to obtain through observations, theoretical calculations give approximate results \citep[e.g.,][]{2015MNRAS.454.1767L} and robust values have been obtained for very few blazars \citep{2005AJ....130.1418J}. 

This has led researchers to come up with various methods to infer the jet power \citep[e.g.,][]{2019hepr.confE..70F}. In many cases the jet power has been calculated using physical parameters of the jet obtained from the modeling of their broadband SED \citep[eg.,][]{2003ApJ...593..667M, 2010MNRAS.402..497G, 2014Natur.515..376G}. Another method is through observations of the pc-scale jet with VLBI, e.g., \cite{2019Ap&SS.364..123C} derive jet power from the observations of the radio core of the jets. Luminosity of extended radio emission from jets have also been shown to correlate with total jet power \citep{2004ApJ...607..800B, 2010ApJ...720.1066C,2013ApJ...767...12G}. Unlike the collimated jet extending to kpc scales, extended emission emanate from slowed plasma and is not significantly variable at the observed timescales. They are free from Doppler boosting and hence provide a better measure of the intrinsic jet power. However, no method is comprehensive because issues like model dependence and lack of a large sample for which a certain method is applicable may distort the results regarding the correlation between disk and jet power. One way to improve on this is to check the correlation obtained from different methods and compare the statistical trend.
\begin{figure*}
\centering
\includegraphics[width=\textwidth]{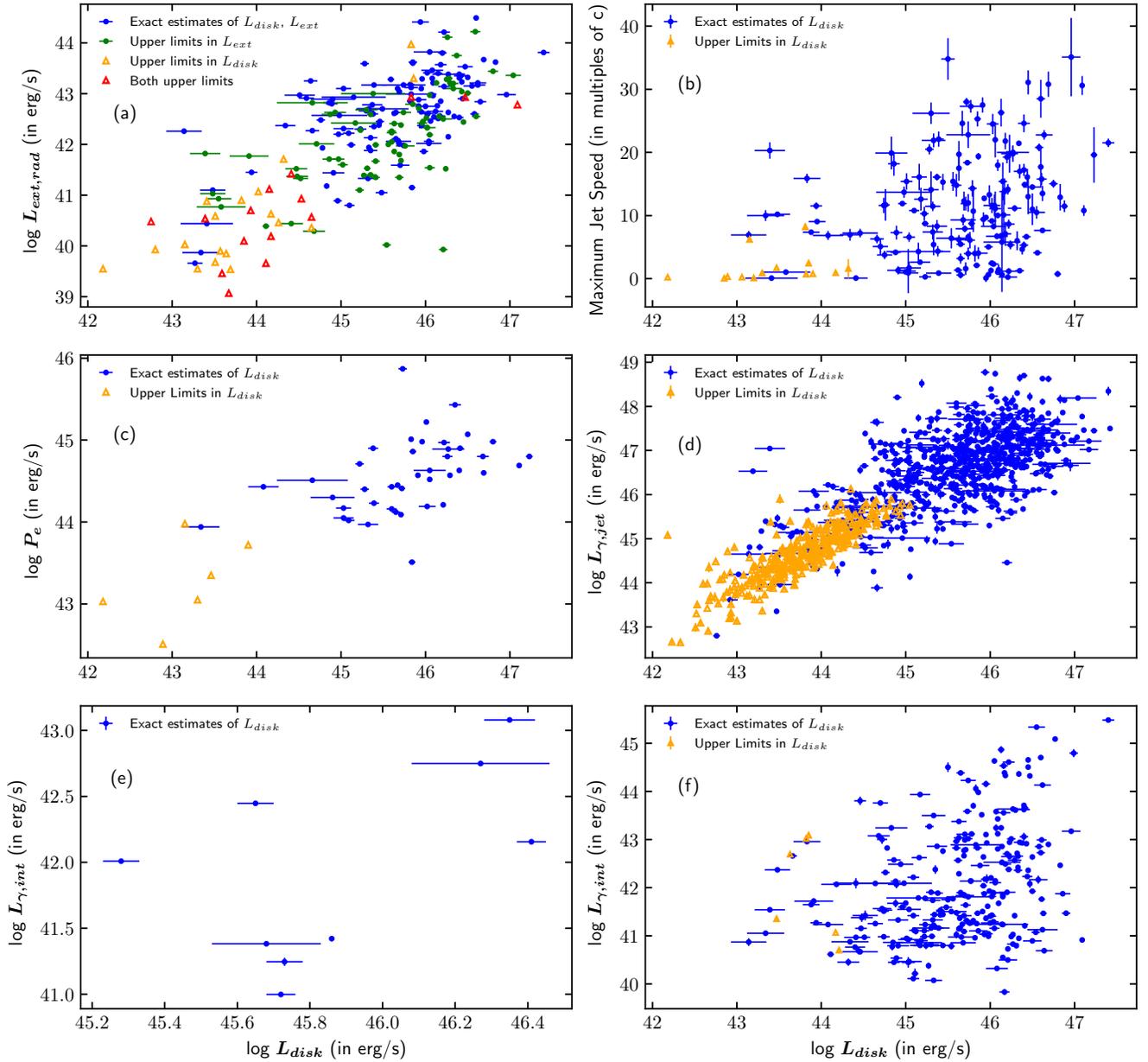}
\caption{The panels show the relation between the disk luminosity of the blazars in our sample with various estimates of jet power. Unless otherwise stated, the blue filled circles represent blazars with observed emission line systems, mainly FSRQs, for which exact estimates of disk luminosity are available in the Vaidehi catalog. The orange open triangles represent sources, mainly BL Lac objects, for which only upper limits to the disk luminosity are available. \textbf{(a)} Relation between the disk luminosity and the low-frequency radio luminosity of the extended jet. The blue and green filled circles represent blazars, mainly FSRQs, for which estimated values and upper limits of the extended radio luminosities are available, respectively, as well as the calculated values of their disk luminosity. The orange and red open triangles consist mainly of BL Lac objects, for which  estimated values and upper limits of the extended radio luminosities are available, respectively, and only the upper limits to their disk luminosity. \textbf{(b)} Relation between the disk luminosity and the superluminal speed in the jet. The correlation is weak although there is no large values of jet speed at smaller values of disk luminosity. \textbf{(c)} Relation between the disk luminosity and the electron kinetic power in jet. \textbf{(d)} Relation between the disk luminosity and the $\gamma$-ray luminosity of jets without beaming correction. \textbf{(e)} Relation between the disk and intrinsic jet luminosity obtained by de-beaming the GeV luminosity with the Doppler factor obtained from VLBA observations. \textbf{(f)} Relation between the disk and intrinsic jet luminosity obtained by de-beaming the GeV luminosity with the variability Doppler factors.} 
\label{fig:panel}
\end{figure*}

In this paper, we venture to study the disk-jet connection in a large sample of blazars by comparing the disk power inferred from the BLR luminosity with five different measures of the jet power and analyzing the trends found using the different indicators. 

In Section \ref{sec2}, we describe our sample and the data used as different indicators of jet power. We compare the disk luminosity with low frequency radio luminosity of the extended jet, jet power inferred from SED modeling, superluminal speeds of radio knots moving down the pc-scale jet observed by VLBA and beamed and debeamed $\gamma$-ray luminosity, in Section \ref{sec3}. In Section \ref{sec4}, we describe the correction for redshift dependence of the luminosity-luminosity correlations shown in Section \ref{sec3}. The results are summarized and their implications are discussed in section \ref{sec5}.

\section{Data}
\label{sec2}
The \textit{Fermi Gamma-ray Space Telescope} has been monitoring the $\gamma$-ray sky since 2008. We use the source name, 0.1-100 GeV energy flux, photon power-law index and blazar class of 1026 blazars from the most recent all-sky catalog of \textit{Fermi} blazars \citep[ 4FGL-DR2 and 4LAC;][]{2020ApJS..247...33A}. The central engine catalog presented in \cite{2021ApJS..253...46P} (hereafter Vaidehi catalog) reports the disk luminosity, redshift, black hole mass and source name of 1077 \textit{Fermi} blazars. In the Vaidehi catalog, the disk luminosities are calculated from emission line luminosities. The blazars have been classified as those with strong and broad emission lines in their spectra and the rest without broad lines. The former group consists mainly of FSRQs although a few BL Lacs show broad lines. The second group mostly contains BL Lacs. We cross-match the sources in the Vaidehi catalog with those in the  4FGL and 4LAC in order to obtain the GeV luminosity of the blazars in our sample. 

We use the luminosity of low-frequency radio emission from \cite{2021MNRAS.505.4726K}, electron kinetic power obtained from \cite{2010MNRAS.402..497G}, and the maximum value of the superluminal speed of the VLBA knots from the MOJAVE data archive \citep{2018ApJS..234...12L} as various measures of jet power. The $\gamma$-ray emission of blazars is relativistically beamed. In order to obtain the intrinsic luminosity of the jets by ``de-beaming,'' we use Doppler factors obtained by \cite{2005AJ....130.1418J} using VLBA monitoring of the pc-scale jet. However, it is available only for 15 blazars, out of which 9 matches with our sample. Therefore, we obtain the variability Doppler factor of a larger fraction of the blazars in our sample from \cite{2018ApJ...866..137L}.

We have used $\Omega_m =0.27$, $\Omega_\Lambda=0.73$, and the redshift given in the Vaidehi catalog to calculate the luminosity from the flux values.

\section{Comparing Disk Luminosity with Various Indicators of Jet Power}
\label{sec3}

\begin{table*}
\caption{\label{tab:table1} For all the samples in Fig. \ref{fig:panel}, the values of correlation coefficients, slope of regression line on log($L_{jet}$) vs log($L_{disk}$) plane and the uncertainty of the slope (as a measure of the scatter of the data) for the sub-sample containing only FSRQs.}
\begin{ruledtabular}
\begin{tabular}{cccc}
 Jet power obtained from&$r$ & Slope\footnote{Slope obtained from linear regression of the log($L_{jet}$)-log($L_{disk}$) data} & Uncertainty of slope\footnote{As a measure of scatter we have listed the uncertainties in the slope \cite{2006NewA...11..567M}. Thus, larger the uncertainty, more the scatter in the correlation.}\\ \hline
Extended Radio Emission&0.630&0.670&0.077 \\
Beamed Gamma ray&0.655&0.722&0.031\\
 De-beamed (variability) Gamma ray \cite{2018ApJ...866..137L}&0.345&0.538&0.091\\
 De-beamed (VLBA) Gamma ray \cite{2005AJ....130.1418J}&0.536&1.024&0.608\\
SED Modeling &0.451&0.267&0.083\\
Speed of pc-Scale Radio Knots&0.258&2.406&1.248\\
\end{tabular}
\end{ruledtabular}
\end{table*}

\subsection{Low-Frequency Radio Luminosity from the Large Scale Jet}
Estimates of the total jet power from the low-frequency (300 MHz) radio luminosity has been provided in \cite{2021MNRAS.505.4726K}, for a sample consisting of over 1000 blazars and radio galaxies. They estimate the 300 MHz flux using spectral decomposition of all available observations at frequencies below $10^{13}$ Hz. Extended low-frequency radio emission arises from slowed plasma in the jets, unlike the collimated superluminal motion seemingly originating from the radio core. Hence, the emission is not Doppler boosted. Since the low-frequency radio jets are extended to hundreds of kpc, there is a time delay between the epochs when the observed jet and disk emission were originally generated. However, the extended jet emission is not variable at timescales relevant for our analysis, thus providing a robust measure of jet power to be compared with the disk luminosity. On cross-matching the above list with the sources in the Vaidehi catalog, we find 221 common blazars. The luminosity of the extended radio emission of these sources are plotted against their disk luminosity (Fig. \ref{fig:panel}(a)).  A strong correlation is indicated by the Pearson coefficient $r=0.793$. Excluding upper limits, we obtain a FSRQ-dominated subset of 115 blazars, which has $r=0.630$ (Blue circles in Fig. \ref{fig:panel}(a)). Table {\ref{tab:table1}} provides the slope of the best-fit straight line and its uncertainty on the log($L_{jet}$) vs log($L_{disk}$) plane. The uncertainty in the slope may be used as a measure of the scatter \mbox{\cite{2006NewA...11..567M}}.

\begin{figure}
\centering
\includegraphics[width=14cm,height=10cm]{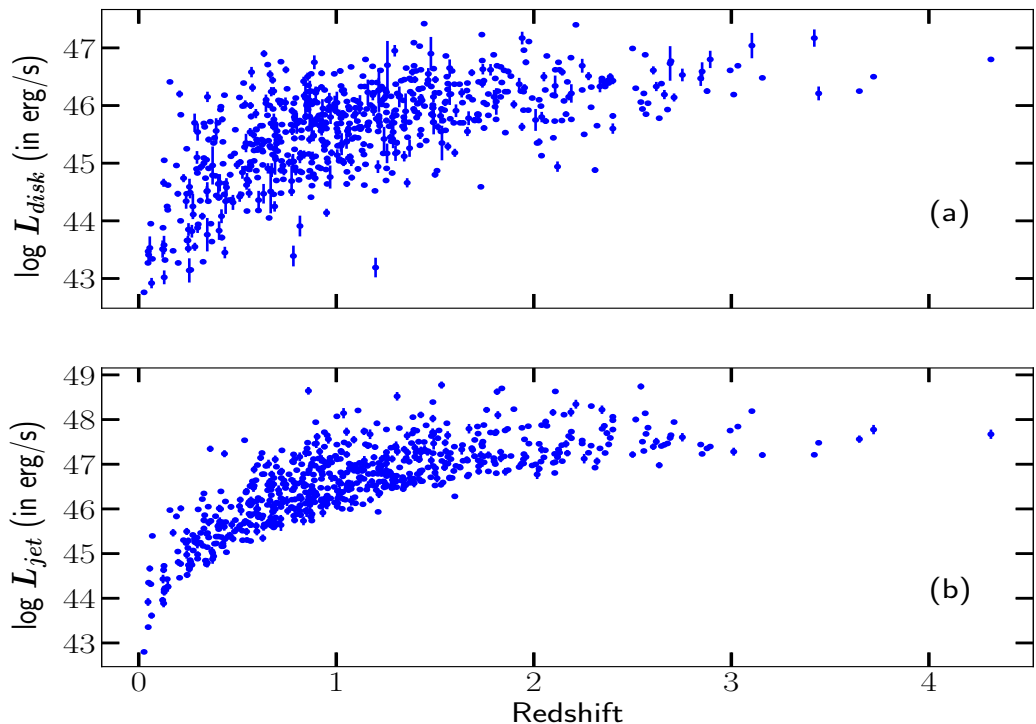}
\caption{Dependence of the disk and jet luminosity of the FSRQ-dominated set on redshift. \textbf{(a)} Blue filled circles represent values of disk luminosity \textit{versus} redshift. \textbf{(b)} Blue filled circles represent values of jet gamma-ray luminosity \textit{versus} redshift.}
\label{fig:diskJet-RedDep}
\end{figure}

\subsection{Speed of pc-Scale Radio Knots from VLBA Monitoring}
The MOJAVE data archive\footnote{\url{https://www.cv.nrao.edu/MOJAVE/}} presents the speed of the radio-bright knots traversing down the pc-scale jet of a large sample of blazars \citep{2018ApJS..234...12L}. The average or maximum speed of such radio knots may provide a suitable estimate of jet power, which is not affected by the uncertainty regarding the radiative efficiency of blazar jets. We find 176 blazars of the Vaidehi catalog for which the speed of the knots are available from the MOJAVE archive. We plot the maximum observed speed of the knots of a given source \textit{versus} the disk luminosity in Fig. \ref{fig:panel}(b). It does not show a strong correlation (Pearson coefficient $r=0.388$). However, the data points are not scattered uniformly on the plane of the plot but are located in a constrained region such that blazars do not exhibit high jet speed for low disk luminosity. On the other hand, there are both high and low jet speeds for high disk luminosity.

\subsection{Jet Power Inferred from SED Modeling}
The kinetic power of the bulk motion of the electrons in the jet is calculated by \cite{2010MNRAS.402..497G} for many blazars. A one-zone leptonic model has been used to fit the SED of each blazar. From the estimated parameters of the fit, the kinetic power of the electron bulk motion has been calculated, which may be used as an indicator of the jet power. We note that the fraction of the jet power that contributes to the kinetic power of the electrons is not well-constrained. Therefore, its value gives a lower limit on the jet power. Its comparison with the disk luminosity is useful in the context of disk-jet relation in a sample of blazars only if the above fraction is approximately constant for all sources. In Fig. \ref{fig:panel}(c), we show the disk luminosity obtained from the Vaidehi catalog along with the jet power determined as above for 48 blazars, which are common to both lists. We compute the Pearson correlation coefficient to be $r=0.741$. For 42 common blazars in the FSRQ-dominated subset, excluding upper limits in $L_{disk}$, we obtain $r=0.451$. 

\subsection{$\gamma-$Ray Luminosity without Beaming Correction}
$\gamma$-ray emission is a suitable indicator of jet power because it is not contaminated by any other part of the AGN. On the other hand, the radiative efficiency of AGN jets is not well constrained, Fermi blazars may be selectively brighter in $\gamma$-rays, and the jet emission in blazars is relativistically beamed, all of which may introduce uncertainties in the jet power computed from the GeV luminosity. However, comparing the disk and $\gamma$-ray emission in a large sample like we have used here may provide an indication whether blazars with stronger disk emission have systematically brighter jets. We find 1026 sources by cross-matching the Vaidehi and Fermi catalogs and we show the result in Fig. \ref{fig:panel}(d). The Pearson correlation coefficient is $r=0.865$ for the entire blazar set, which indicates that the disk and $\gamma$-ray luminosity are strongly correlated. For a subset of the parent population, containing mainly the BL Lac objects, the exact value of the disk luminosity is not available and only upper limits are given. Excluding those sources, for a subset of our sample dominated by FSRQs (684 sources; shown as blue filled circles in Fig. \ref{fig:panel}(d)), we found the correlation coefficient to be $r=0.656$. However, both disk and jet luminosities depend on the redshift, and the intrinsic correlation may be weaker. We discuss this in Section \ref{sec4}.
\begin{figure}
\centering
\includegraphics[width=14cm,height=10cm]{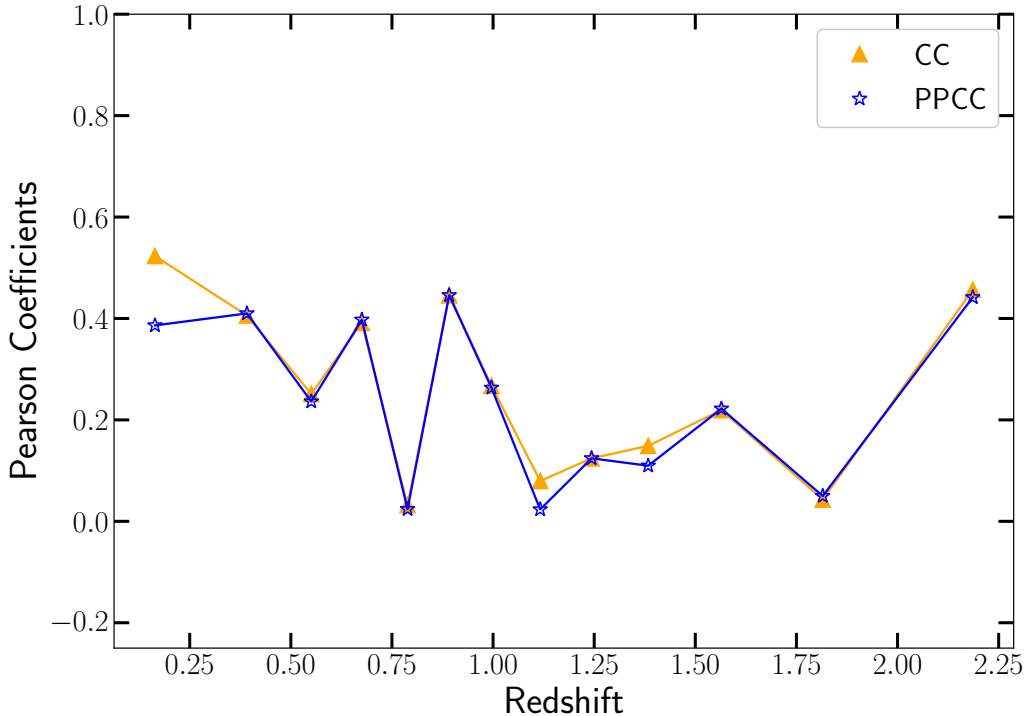}
\caption{Values of Pearson's coefficients \textit{versus} redshift of the FSRQ-dominated set. PPCC denotes partial correlation and CC denotes simple correlation coefficient. The correlation is weak except at the lowest and highest ends of the redshift range.}
\label{fig:RedBin}
\end{figure}

\subsection{$\gamma$-ray Luminosity with Beaming Correction}
Blazar emission is relativistically beamed in the observer's frame. The observed jet luminosity ($L_\gamma$) may be written in terms of the intrinsic luminosity ($L_{\gamma,int}$) and the Doppler beaming factor $\delta$ as \citep{2012ApJ...751..108A},
\bbe
L_\gamma &=& \delta^p L_{\gamma,int}
\ee
where, the value of $p$ is between 4 and 6, depending on the emission mechanism, and the dynamics of the jet. We use $p=4$. Therefore, to obtain the intrinsic luminosity of the blazar jet the value of $\delta$ needs to be known. Below, we describe two methods by which the value of $\delta$ may be estimated and the relation between the disk and the beaming-corrected intrinsic jet luminosity in the GeV band that we obtain using those estimates.

\subsubsection{Doppler Factor Inferred from VLBA Monitoring}
The value of $\delta$ for several blazars has been estimated in \cite{2005AJ....130.1418J}, using monthly Very Long Baseline Array (VLBA) monitoring of the pc-scale jet of a sample of blazars. However, this method is observing time intensive and may be carried out only for a small sample. On cross-matching with their list, we find nine blazars of our sample for which the value of $\delta$ is available. We correct the observed $\gamma$-ray luminosity using those values and show their relation with the disk emission in Fig. \ref{fig:panel}(e). We calculate the Pearson correlation coefficient to be $r=0.537$.

\subsubsection{Variability Doppler Factor}
The value of $\delta$ for 1029 blazars and radio galaxies has been presented by \cite{2018ApJ...866..137L}, using the 15 GHz radio monitoring program with the 40 m telescope at the Owens Valley Radio Observatory. This method was found to be more accurate than other recent attempts at constraining $\delta$, using blazar population models \cite{2015MNRAS.454.1767L}. Variability Doppler factors \citep{1999ApJ...521..493L,2009A&A...494..527H} are estimated by comparing the intrinsic equipartition temperature associated with the jet to the highest observed brightness temperature. The intrinsic brightness temperature is obtained at the peak of the prominent jet flares \citep{1999ApJ...521..493L}, using the assumption of equipartition of energy density of the magnetic field and the radiating particle \citep{1994ApJ...426...51R}. On cross-matching, we find 265 blazars of our sample for which the value of $\delta$ is available. We correct the observed $\gamma$-ray luminosity using those values and show their relation with the disk emission in Fig. \ref{fig:panel}(f). We calculate the Pearson correlation coefficient to be $r=0.345$. We find that the points are scattered in a pattern similar to Fig. \ref{fig:panel}(b), i.e., the maximum speed of radio knots in the jet \textit{versus} disk luminosity. We discuss this similarity further in Section \ref{sec4}.

\begin{figure}
\centering
\includegraphics[width=14cm,height=10cm]{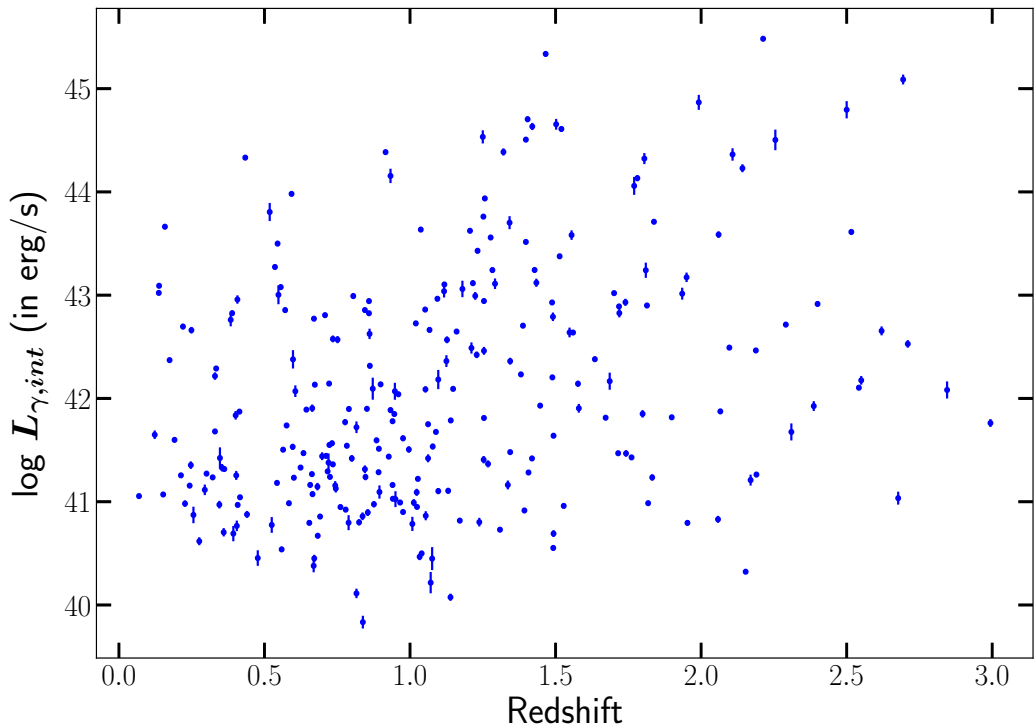}
\caption{Relation between the de-beamed GeV luminosity and redshift.}
\label{fig:RedIntJet}
\end{figure}

\begin{figure*}
\centering
\includegraphics[width=\textwidth]{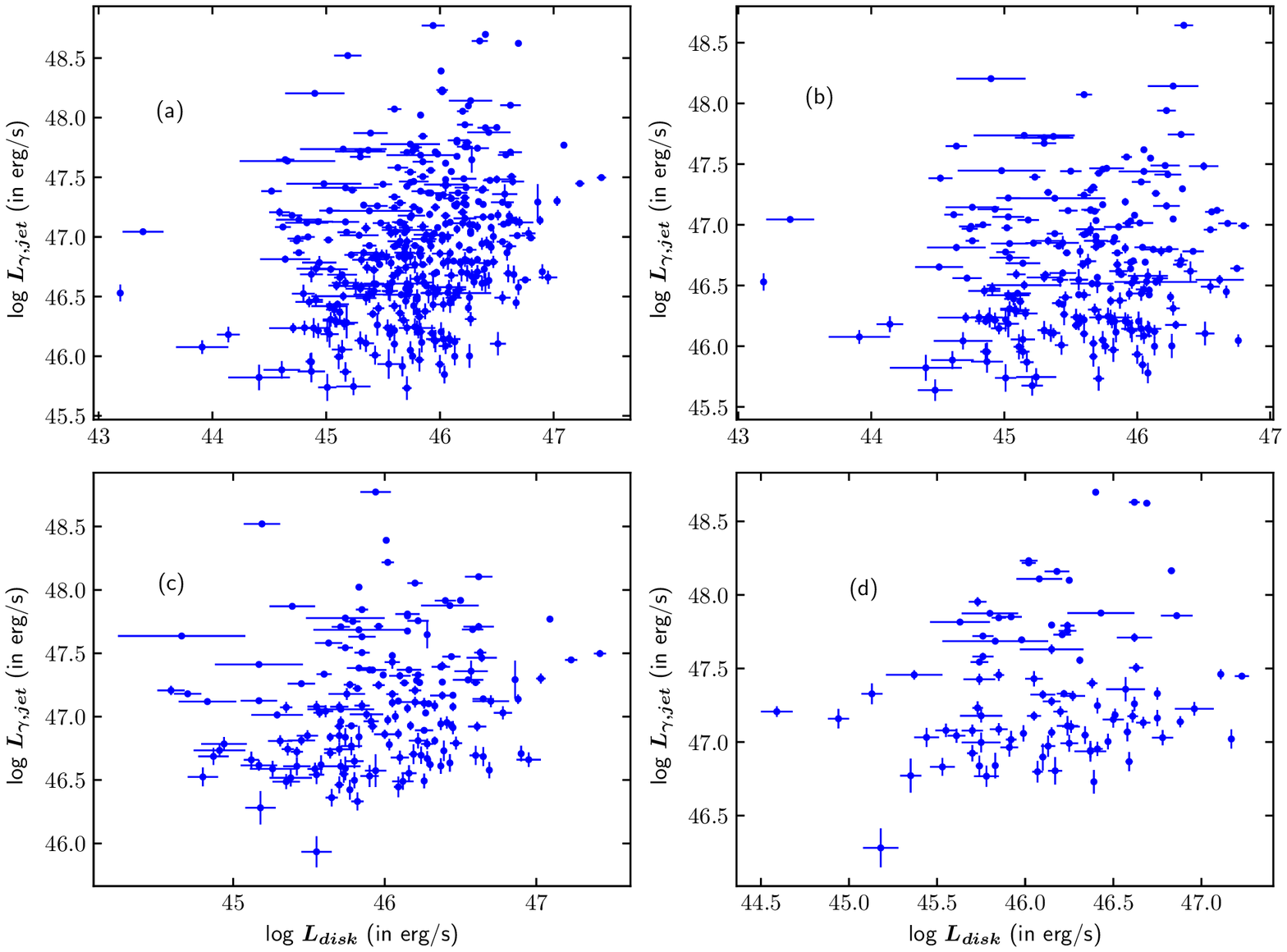}
\caption{Blue solid circles with error bars demonstrate the relation between the beamed GeV luminosity and disk luminosity of sources in various redshift intervals for the FSRQ-dominated set. \textbf{(a)} Redshift interval of 0.78 to 1.9. \textbf{(b)} Redshift interval of 0.7 to 1.2 \textbf{(c)} Redshift interval of 1.2 to 1.8 \textbf{(d)} Redshift interval of 1.6 to 2.2. In all of the cases, the correlation is weak although there is no large values of jet luminosity at smaller values of disk luminosity.}
\label{fig:RedInt}
\end{figure*}

\section{Correction for Redshift dependence of Luminosity-Luminosity Correlation}
\label{sec4}
In the previous section we have shown the relation of the disk luminosity of a large sample of blazars with a number of quantities, which are indicators of jet power. Among those, besides the speed of radio knots obtained from VLBA monitoring, all are luminosities observed or calculated in different methods. Most of those quantities show a significant correlation with the disk luminosity. However, it is well known that any correlation between a pair of luminosities are often driven by their dependence on redshift or distance rather than their intrinsic properties. In a magnitude limited survey, objects that are farther away tend to be brighter due to selection effect. On the other hand, objects that are nearer tend to be fainter because in a small volume, fraction of faint objects is higher. Consequently, in this case, farther objects will tend to have larger values of disk as well as jet luminosity while nearer objects will have smaller values of both. That will cause the sample to seem apparently correlated  \citep{1983ApJ...275...45C,1983ApJ...269..400F,1986MNRAS.220...51K,1996MNRAS.278..919A}.

To test the reliability of the correlations obtained in the previous section, we check the dependence of the luminosities on redshift. We plot the $\gamma$-ray luminosity without beaming correction and the disk luminosity of the FSRQ-dominated subset with redshift in Fig. \ref{fig:diskJet-RedDep}. We find that both of those luminosities are strongly dependent on redshift.

We use two methods to obtain the intrinsic correlation between the pairs of luminosities: Pearson's partial correlation coefficient (PPCC) and calculation of the correlation coefficient in smaller redshift bins. Pearson's partial correlation coefficient (PPCC) is used to measure the correlation between two quantities, subtracting their common dependence on a third variable. In our case, disk and jet luminosities are the relevant quantities while we need to remove the common redshift ($z$) dependence. PPCC is defined as \citep[e.g.,][]{1992A&A...256..399P}
\bbe
r_{XY,Z} &=& \frac{r_{XY}-r_{XZ}r_{YZ}}{\sqrt{(1-r^{2}_{XZ})(1-r^{2}_{YZ})}},
\ee
where $r_{XY,Z}$ is the partial correlation of $X$ and $Y$ discounting for $Z$, and $r_{XY}$ is Pearson's simple correlation coefficient (CC) between $X$ and $Y$. We calculate the PPCC to mathematically describe the strength of partial correlation among the disk luminosity ($L_D$) and jet luminosity ($L_j$) in our sample. 

We obtain a partial correlation of $r_{L_{d}L_{\gamma},z} = 0.382$ on using the FSRQ-dominated set, which is significantly less than the CC value for the same set, in $\gamma$-ray luminosity. The correlation of the disk luminosity ($L_D$) with the low-frequency radio ($L_{ext}$) may be similarly affected by the redshift dependence of the luminosities. Therefore, we calculate the PPCC and find, $r_{L_{d}L_{ext},z} = 0.338$ for the FSRQ dominated subset. The value of the PPCC is significantly smaller than the CC values found in the previous section. This indicates that the disk-jet correlation is much weaker when the redshift dependence is removed.

Low frequency radio emission was used by \cite{fan2019jet}, from the TIFR Giant metrewave radio telescope Sky Survey (TGSS) catalog at 150 MHz \citep{IntemaTGSS2017} and bolometric luminosity from the literature as the measure of jet and disk power, respectively, to probe the disk-jet connection. They found a weak correlation with $\tau = 0.28$ on using partial Kendall's $\tau$ correlation test after removing the redshift bias, which is consistent with the results we obtain. On extending the partial correlation analysis to our entire blazar set (1026 sources), we get $r_{L_{d}L_{\gamma},z} = 0.634$. This is consistent with the results presented in \cite{2014MNRAS.445...81S}, who used a sample of FSRQs, BL Lacs and radio galaxies and analyzed $L_{BLR}-L_\gamma$ to obtain PPCC = 0.59 and $L_{BLR}-L_{radio}$ to obtain PPCC = 0.65. 

Next we divide the data into smaller bins in redshift, each containing 50 sources. We have kept the number of sources fixed for each bin so that the correlation coefficients obtained for each bin may be meaningfully compared. This is a useful method to probe intrinsic correlation in luminosity-luminosity data  \citep[e.g.,][]{2019ApJ...877...63S}.  Obviously, the redshift interval for each bin is not the same in order to keep the number of sources constant. To determine the fixed number of sources to be contained in each bin, we choose the highest possible number of sources for which PPCC and CC are approximately equal in the individual intervals. This is because redshift binning is carried out to remove the effect of redshift bias. Thus PPCC and CC are expected to be of similar value for the individual bins. After optimizing the number of sources per bin to be 50, we plot the CC and PPCC of the $\gamma$-ray and disk luminosities of the FSRQ-dominated blazar set in each bin in Fig. \ref{fig:RedBin}. It is evident that the correlation coefficients are low at the middle of the redshift range and the higher values are located at the highest and lowest ends of the redshift range of our sample.

To understand the redshift dependence of the de-beamed GeV luminosities of the jets, we plot the intrinsic $\gamma$-ray luminosity vs redshift in Fig. \ref{fig:RedIntJet}. We find that, unlike the quantities shown in Fig. \ref{fig:diskJet-RedDep}, there is no correlation between the de-beamed GeV luminosity with redshift. The Doppler factors obtained from variability are redshift-independent and the intrinsic $\gamma$-ray luminosity calculated by de-beaming with those Doppler factors have resulted in the redshift-independent nature of $L_{\gamma,int}$ in this case. We note that the pattern of scatter in Fig. \ref{fig:panel}(b) and Fig. \ref{fig:panel}(f) are similar. Both quantities, namely, apparent speed of pc-scale radio knots and $L_{\gamma,int}$ are redshift-independent and they show a very similar pattern.

Since we see that the major decrease in the correlation coefficient occurs in the redshift interval $z = 0.78$ to 1.9, we plot the disk and jet luminosities for the blazars in this interval separately in Fig. \ref{fig:RedInt}(a). We can clearly see that the correlation is much weaker than what we find for the entire sample in Figure \ref{fig:panel}(d). This is consistent with the notion that the correlation we see when we consider the entire sample is driven by the high and low end of the redshift ranges. However, we note that while there is scatter in the plot there is no high value of jet luminosity for low value of disk luminosity, although there are sources having both high and low jet luminosity for high disk luminosity. This constrains the scatter plot to a particular side of the $L_{disk}-L_{jet}$ plane. In Fig. {\ref{fig:RedInt}} we further divide the redshift interval into smaller bins and study the scatter of objects belonging to each bin. Fig. {\ref{fig:RedInt}} (b),(c) and (d) are subsets of Fig. {\ref{fig:RedInt}} (a) belonging to the redshift interval indicated in the figure captions. We find that in each case of smaller redshift bins, the distribution of data points remains approximately constrained in one part of the $L_{disk}-L_{jet}$ plane as is the case of Figure \ref{fig:panel}(b) and \ref{fig:panel}(f). That is consistent with the notion that while higher jet powers are always accompanied by higher disk luminosities, objects with low jet powers may possess a range of disk powers from low to high.

Finally, we carry out the so-called ``scrambling test'' \cite{2005astro.ph.11368B,2006NewA...11..567M,2009A&A...501..915B} to determine the probability that the correlation we obtain among different pairs of quantities is by chance due to the underlying common redshift dependence. 
In this method, we keep the disk luminosity fixed along with the corresponding redshifts. We assign jet flux (we use two cases: extended radio and beamed gamma-ray fluxes) to the disk luminosities in random permutations. We calculate the jet luminosities using the newly assigned redshifts. This includes the effect of distance, removing any intrinsic physical correlation in the data set since the jet fluxes are scrambled. To estimate the chance probability, we perform a Monte Carlo test by repeating the above method $10^4$ times, creating a new randomized data set each time. By comparing the correlation coefficient we obtained from the actual data with the distribution of the same from the scrambled data sets described above, we find that for the case of the disk luminosity vs extended radio luminosity, there is a $\sim$3\% probability of obtaining a similar or stronger correlation by chance, i.e., due to the common redshift dependence. Therefore, the intrinsic correlation in that case is weak. In the case of disk luminosity vs beamed gamma-ray luminosity the corresponding probability is negligibly small indicating that correlation is stronger.

\section{Summary and Discussion}
\label{sec5}
We have studied the accretion disk versus jet power in a large sample of Fermi blazars. The disk luminosities have been obtained from literature, and are inferred from the luminosity of the emission lines. Intrinsic jet power, on the other hand, has no unique well established observational indicator. Therefore, five different quantities have been used as measures of jet power in our analysis. At first, we studied the correlation between the disk luminosity and low frequency radio luminosity of the extended jet, which is presumably unbeamed and not variable at the relevant timescales in this case. We found a significant correlation between those two quantities. Next we used other indicators of jet power, such as, electron kinetic power in the jet obtained using the best-fit parameters from SED modeling and beamed $\gamma$-ray luminosity. In all of those cases, we obtained a significant correlation between the disk and jet power. However, we found that those correlations are partially driven by the redshift dependence of the pair of luminosities being correlated in each case. Upon calculating the partial correlation, in which the common redshift dependence of the pairs of quantities is removed, we obtained a weaker correlation in each of the above cases. A weak correlation is also obtained on using the debeamed GeV luminosity as jet power, which has been shown to have almost no redshift dependence. Finally, we used the speed of the radio-bright knots moving down the pc-scale jet as observed with the VLBA, as an indicator of the jet power. This quantity is not redshift dependent and shows a weak correlation with disk luminosities.

Figure \ref{fig:RedInt} shows the disk-jet scatter in redshift intervals, in which the redshift bias is shown to be irrelevant (in Fig. \ref{fig:RedBin}). This pattern is remarkably similar to Fig. \ref{fig:panel}(b) and Fig. \ref{fig:panel}(f), in which the apparent speed of superluminal knots and beaming-corrected GeV luminosity have been used as indicators of jet power, both of which happen to be independent of redshift. This may imply that a powerful disk is a necessary but not sufficient condition to generate a powerful jet. Other parameters, such as BH spin, environment of the jet and its interaction with it may affect the observed power of the jet. We conclude that the disk and jet power are indeed correlated but the correlation is weakened by scatter caused due to the above factors.

\section*{Data Availability}

The data utilized in this work have been obtained from:-
\begin{enumerate}
    \item The central engine catalog presented in \cite{2021ApJS..253...46P}.\\ Weblink: \url{http://www.ucm.es/blazars/engines}
    \item Fermi 4FGL-DR2 and 4LAC catalogs.\\ Weblink: \url{https://fermi.gsfc.nasa.gov/ssc/data/access/lat/10yr_catalog/}
    \item Luminosity of extended radio emission has been collected from \cite{2021MNRAS.505.4726K} as a measure of total jet luminosity.
    \item The gamma ray beaming factors for 15 blazars are found in \cite{2005AJ....130.1418J}. 
    \item Variability Doppler factors have been obtained from \cite{2018ApJ...866..137L}
    \item Electron kinetic power has been obtained from \cite{2010MNRAS.402..497G} as a measure of jet power.
    \item The superluminal speed in jets has been gathered from the MOJAVE data archive and used as a measure of intrinsic kinetic power of jet \cite{2018ApJS..234...12L}.\\ Weblink: \url{https://www.cv.nrao.edu/MOJAVE/}
\end{enumerate}
\vspace{1.5mm}

\begin{acknowledgments}
We thank the anonymous referee for comments that helped in improving the manuscript. GR acknowledges the DST INSPIRE Scholarship and JBNSTS Scholarship. RC thanks Presidency University for support under the Faculty Research and Professional Development (FRPDF) Grant, ISRO for support under the AstroSat archival data utilization program, and IUCAA for their hospitality and usage of their facilities during his stay at different times as part of the university associateship program. RC acknowledges financial support from BRNS through a project grant (sanction no: 57/14/10/2019-BRNS) and thanks the project coordinator Pratik Majumdar for support regarding the BRNS project. 
\end{acknowledgments}



%

\end{document}